\documentclass[twocolumn,amsmath,showpacs,amsfonts,aps,prc,floatfix]{revtex4}
\usepackage{graphicx}
\usepackage{bm}
\newcommand{\dn}{\frac{dN^\phi}{dy}} 
\newcommand{\pt}{<p_T^\phi>} 
\begin{document}

\title{Nearly perfect fluid in Au+Au collisions at RHIC}
 
\author{A. K. Chaudhuri}
\email[E-mail:]{akc@veccal.ernet.in}
\affiliation{Variable Energy Cyclotron Centre, 1/AF, Bidhan Nagar, 
Kolkata 700~064, India}

\begin{abstract}

In the Israel-Stewart's theory of dissipative hydrodynamics,   
we have analysed the STAR data on $\phi$ meson production in Au+Au collisions at $\sqrt{s}$=200 GeV. From a simultaneous fit to
$\phi$ mesons multiplicity, mean $p_T$ and integrated $v_2$, we obtain a phenomenological estimate of QGP viscosity, $\eta/s =0.07 \pm 0.03 \pm 0.14$,
the first error is due to the experimental uncertainty in STAR measurements, the second reflects the uncertainties in initial and final conditions of the fluid.


 \end{abstract}


\date{\today}  

\maketitle


 
In recent years, there is considerable interest in viscosity of 
strongly interacting Quark-Gluon Plasma.
String theory based models (ADS/CFT) give a lower bound on viscosity of any matter $\eta/s \geq 1/4\pi$ \cite{Policastro:2001yc}. In a perturbative QCD, Arnold et al  \cite{Arnold:2000dr} estimated $\eta/s\sim$ 1. 
In a SU(3) gauge theory, Meyer \cite{Meyer:2007ic} gave the upper bound $\eta/s <$1.0, and his best estimate is $\eta/s$=0.134(33) at $T=1.165T_c$. 
At RHIC region, Nakamura and Sakai \cite{Nakamura:2005yf}
estimated the viscosity of a hot gluon gas  as $\eta/s$=0.1-0.4. Attempts have been made to estimate QGP viscosity directly from experimental data. 
Gavin and Abdel-Aziz \cite{Gavin:2006xd} proposed to measure viscosity from transverse momentum fluctuations. From the existing data on Au+Au collisions, they estimated that QGP viscosity as $\eta/s$=0.08-0.30. Experimental data on elliptic flow has also been used to estimate QGP viscosity. Elliptic flow scales with eccentricity. Departure form the scaling can be understood as due to off-equilibrium effect and utilised to estimate viscosity \cite{Drescher:2007cd} as, $\eta/s$=0.11-0.19. Experimental observation that elliptic flow scales with transverse kinetic energy is also used to estimate QGP viscosity, $\eta/s \sim$ 0.09 $\pm$ 0.015 \cite{Lacey:2006bc}, a value close to the ADS/CFT bound. From heavy quark energy loss, PHENIX collaboration \cite{Adare:2006nq} estimated
QGP viscosity $\eta/s\approx$ 0.1-0.16.  Recently, from analysis of RHIC data,
in a viscous hydrodynamics, upper bound to viscosity is given $\eta/s <$ 0.5 \cite{Luzum:2008cw,Song:2008hj}.

In the present paper, from a hydrodynamic analysis of the recently measured STAR data \cite{Abelev:2007rw} on $\phi$ production, we have obtained an accurate estimate of QGP viscosity, $\eta/s$=0.07 $\pm$ 0.03 $\pm$ 0.14, the first error corresponding the uncertainty in STAR measurements, the 2nd error arising from the uncertain initial conditions of the fluid, e.g. initial time, initial fluid velocity, freeze-out temperature etc. As noted in \cite{Mohanty:2009tz}, several unique features of $\phi$ mesons (e.g.
hidden strange particle, both hadronic and leptonic decay, not affected by resonance decays,   mass and width are not modified in a medium \cite{Alt:2008iv}  etc.) make it an ideal probe to investigate medium properties in heavy ion collisions. 
For long, strangeness enhancement is considered as a signature of QGP formation \cite {Koch:1986ud}. Compared to a hadron gas, in QGP, 
strangeness is enhanced due to abundant $gg\rightarrow s\bar{s}$ reactions. Early produced $s\bar{s}$, if survive
hadronisation can lead to increased production of strange particles compared to pp or pA collisions. Experimental data do show
strangeness enhancement \cite{strange}. In STAR measurements \cite{Abelev:2007rw} also, compared to a pp collision, in a Au+Au collision, $\phi$ meson production is enhanced. STAR data \cite{Abelev:2007rw} also appear to be consistent with a model based on recombination of thermal strange quarks \cite{Hwa:2006vb}. As it will be shown below, STAR data on $\phi$ mesons are also consistent with hydrodynamic model and are sensitive enough to give an accurate estimate of QGP   viscosity.
  
Space-time evolution of the QGP fluid is obtained by solving, Israel-Stewart's theory of 2nd order dissipative hydrodynamics. 
 
\begin{eqnarray}  
\partial_\mu T^{\mu\nu} & = & 0,  \label{eq3} \\
D\pi^{\mu\nu} & = & -\frac{1}{\tau_\pi} (\pi^{\mu\nu}-2\eta \nabla^{<\mu} u^{\nu>}) \nonumber \\
&-&[u^\mu\pi^{\nu\lambda}+u^\nu\pi^{\nu\lambda}]Du_\lambda. \label{eq4}
\end{eqnarray}

Eq.\ref{eq3} is the conservation equation for the energy-momentum tensor, $T^{\mu\nu}=(\varepsilon+p)u^\mu u^\nu - pg^{\mu\nu}+\pi^{\mu\nu}$, 
$\varepsilon$, $p$ and $u$ being the energy density, pressure and fluid velocity respectively. $\pi^{\mu\nu}$ is the shear stress tensor (we have neglected bulk viscosity and heat conduction). Eq.\ref{eq4} is the relaxation equation for the shear stress tensor $\pi^{\mu\nu}$.   
In Eq.\ref{eq4}, $D=u^\mu \partial_\mu$ is the convective time derivative, $\nabla^{<\mu} u^{\nu>}= \frac{1}{2}(\nabla^\mu u^\nu + \nabla^\nu u^\mu)-\frac{1}{3}  
(\partial . u) (g^{\mu\nu}-u^\mu u^\nu)$ is a symmetric traceless tensor. $\eta$ is the shear viscosity and $\tau_\pi$ is the relaxation time.  It may be mentioned that in a conformally symmetric fluid relaxation equation can contain additional terms  \cite{Song:2008si}.

\begin{figure}[t]
\vspace{0.3cm} 
\center
 \resizebox{0.35\textwidth}{!}{%
  \includegraphics{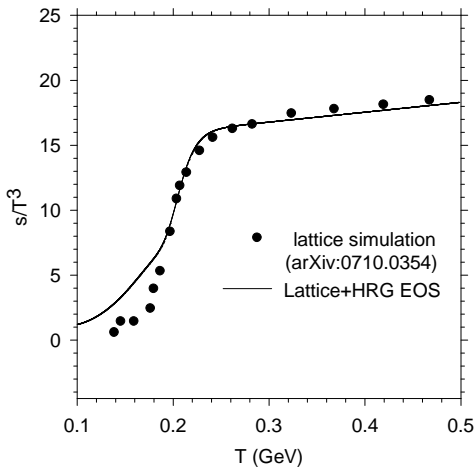}
}
\caption{Black circles are lattice simulation \cite{Cheng:2007jq} for entropy density. The black line is the model EOS, obtained by parametric representation to the lattice simulations and a hadronic resonance gas at low temperature.
}
  \label{F1}
\end{figure}

\begin{figure}[t]
 \vspace{0.3cm} 
 \center
 \resizebox{0.35\textwidth}{!}{%
  \includegraphics{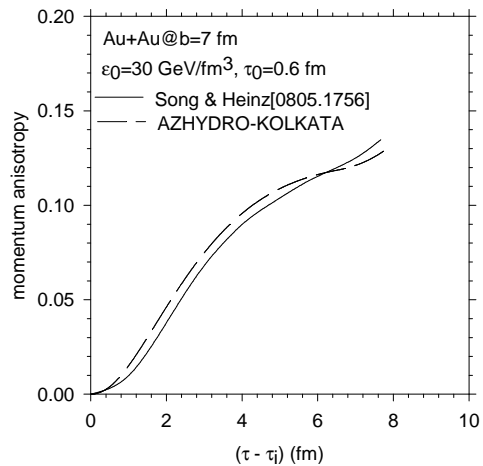}
}
\caption{Viscous fluid ($\eta/s$=0.08) simulation for temporal evolution of momentum anisotropy in b=7 fm Au+Au collision at RHIC. The solid line is the simulation result from VISH2+1 \cite{Song:2008si} 
and the dashed line is the simulation result from AZHYDRO-KOLKATA.}\label{F2}
\end{figure}

Eqs.\ref{eq3},\ref{eq4} are closed with an equation of state $p=p(\varepsilon)$.
Lattice simulations \cite{lattice,Cheng:2007jq} indicate that the confinement-deconfinement transition is a cross over, rather than a 1st or 2nd order phase transition.   In Fig.\ref{F1},  a recent lattice simulation  \cite{Cheng:2007jq} for the  entropy density is
  shown. We complement the lattice simulated EOS \cite{Cheng:2007jq} by a
  hadronic resonance gas (HRG) EOS comprising all the resonances below mass 2.5 GeV. In Fig.\ref{F2}, the solid line is the
   entropy density of the "`lattice +HRG"' EOS. The entropy density is obtained as,
     
   \begin{equation}
   s=0.5[1-tanh(x)]s_{HRG} + 0.5 [1-tanh(x)]s_{lattice}
   \end{equation}
   
\noindent   with $x=\frac{T-T_c}{\Delta T}$, $T_c$=196 MeV, $\Delta T=0.1T_c=19.6 MeV$.  Compared to lattice simulation, entropy density drops slowly at low temperature. 
   
Assuming boost-invariance, Eqs.\ref{eq3} and \ref{eq4}  are solved in $(\tau=\sqrt{t^2-z^2},x,y,\eta_s=\frac{1}{2}\ln\frac{t+z}{t-z})$ coordinates, with a code 
  "`AZHYDRO-KOLKATA"', developed at the Cyclotron Centre, Kolkata.
 Details of the code can be found in \cite{Chaudhuri:2008sj}. To show that AZHYDRO-KOLKATA computes the evolution correctly, in Fig.\ref{F1}, we have compared the temporal evolution 
 of   momentum anisotropy $\varepsilon_p=\frac{<T^{xx}-T^{yy}>}{<T^{xx}+T^{yy}>}$ of a QGP fluid with a calculation
 of Song and Heinz \cite{Song:2008si}. Initial conditions are approximately same in both the simulations.
Within 10\% or less, AZHYDRO-KOLKATA simulation  reproduces  Song and Heinz's  \cite{Song:2008si} result for temporal evolution of momentum anisotropy $\varepsilon_p$.

Solution of partial differential equations (Eqs.\ref{eq3},\ref{eq4}) requires initial conditions, e.g.  transverse profile of the energy density ($\varepsilon(x,y)$), fluid velocity ($v_x(x,y),v_y(x,y)$) and shear stress tensor ($\pi^{\mu\nu}(x,y)$) at the initial time $\tau_i$. One also need to specify the viscosity ($\eta$) and the relaxation time ($\tau_\pi$). A freeze-out prescription is also needed to convert the information about fluid energy density and velocity to particle spectra and compare with experiment.

We assumed that the fluid is thermalised at $\tau_i$=0.6 fm \cite{QGP3} and the initial fluid velocity is zero, $v_x(x,y)=v_y(x,y)=0$.   Initial energy density is assumed to be distributed as \cite{QGP3}

\begin{equation} \label{eq6}
\varepsilon({\bf b},x,y)=\varepsilon_i[0.75 N_{part}({\bf b},x,y) +0.25 N_{coll}({\bf b},x,y)],
\end{equation}

\noindent
where b is the impact parameter of the collision. $N_{part}$ and $N_{coll}$ are the transverse profile of the average number of  participants and average number collisions respectively.
$\varepsilon_i$ is a parameter which does not depend on the impact parameter of the collision. As will be discussed below, we fix it to reproduce experimental data on $\phi$ mesons. Finally, the freeze-out was fixed at $T_F$=150 MeV \cite{note1}. The shear stress tensor was initialised with boost-invariant value, $\pi^{xx}=\pi^{yy}=2\eta/3\tau_i$, $\pi^{xy}$=0. For the relaxation time, we used the   Boltzmann estimate $\tau_\pi=3\eta/2p$.

 \begin{figure}[t]
\vspace{0.3cm} 
\center
 \resizebox{0.35\textwidth}{!}{%
  \includegraphics{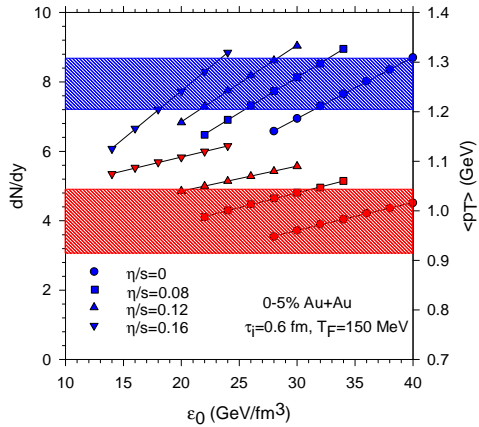}
}
\caption{(color online) 
Variation of $\dn$ and $\pt$ with central energy density in b=2.3 fm Au+Au collisions in AZHYDRO-KOLKATA.
  The blue and red symbols are the predicted $\dn$ and $\pt$ respectively for $\eta/s$=0.0, 0.08, 0.16, 0.2 and 0.25. The blue and red shaded regions indicate the STAR measurements (statistical and systematic error included) of $\dn$ and $\pt$ in 0-5\% centrality Au+Au collisions. } \label{F3}
\end{figure}

In Fig.\ref{F3}, for   fluid viscosity, $\eta/s$=0, 0.08, 0.12 and 0.16, hydrodynamic predictions for $\phi$ meson multiplicity ($\dn$) and mean $p_T$ ($\pt$), in b=2.3 fm Au+Au collisions are shown as  a function of central energy density.   In a hydrodynamic model, as expected, both $\dn$ and $\pt$ increases with increasing energy density, but increase in $\dn$ is steeper than in $\pt$.
b=2.3 fm Au+Au collisions corresponds to 0-5\% centrality collisions. In Fig.\ref{F3}, the blue and red shaded regions represent the STAR measurements (statistical and systematic error  included) on $\phi$ mesons multiplicity and mean $p_T$ in 0-5\% Au+Au collisions. 
One note that irrespective of fluid viscosity, $\phi$ meson multiplicity can be fitted in the hydrodynamic model by changing the initial energy density, more viscous fluid requiring less energy density. It is understood. In viscous fluid evolution, entropy is generated and fluid requires less initial entropy density or energy density. 
However,   with increasing $\eta/s$, mean $p_T$ also increases, even if, $\phi$ meson multiplicity is kept fixed by reducing initial energy density. The reason is understood. Initial transverse pressure increases with increasing $\eta/s$  leading to increased $<p_T>$.   
One note that  simultaneous fit to STAR data on $\phi$ multiplicity and mean $p_T$ in 0-5\% centrality Au+Au collisions is obtained only when $\eta/s \leq 0.12$.   For higher viscosity, while it is possible to fit $\phi$ meson multiplicity, mean $p_T$ is over predicted.


  \begin{table}[ht]
\caption{\label{table1} The fitted values of the initial central energy density ($\varepsilon_i$) and temperature ($T_i$) of the fluid in b=0 Au+Au collisions, for different values of viscosity to entropy ratio ($\eta/s$). In the last four rows, $\chi^2/N$ for STAR data on $\frac{dN^\phi}{dy}$,  $<p_T^\phi>$,$v_2$
and the combined data sets ($\frac{dN^\phi}{dy}$+$<p_T^\phi>$+$v_2$) are shown.} 
\begin{ruledtabular} 
  \begin{tabular}{|c|c|c|c|c|}\hline
$\eta/s$         & 0    & 0.08 & 0.12 & 0.16 \\  \hline
$\varepsilon_i (GeV/fm^3)$ & $35.5$ & $29.1$ & $25.6$ &  $20.8$ \\  
  & $\pm$ 5.0 & $\pm$ 3.6 & $\pm$ 4.0 &  $\pm$ 2.7  \\ \hline
$T_i$ (MeV) & 377.0 & 359.1 & 348.0 & 330.5     \\ 
  & $\pm 13.7$ & $\pm 11.5$ & $\pm 14.3$ & $\pm 11.3$   \\ \hline  
$\chi^2/N (dN/dy)$ & 4.3 & 4.5 &  3.9 & 2.21\\ \hline
$\chi^2/N (<p_T>)$ & 0.55 & 0.26 &  1.80 & 6.22\\ \hline
$\chi^2/N (v_2)$ & 4.92 & 2.99&  2.79 & 3.03 \\ \hline
$\chi^2/N $ &  &  &  & \\  
(dN/dy+ $<p_T>$+$v_2$)& 9.77 & 7.76 & 8.49 & 11.46\\
\end{tabular}\end{ruledtabular}  
\end{table}

\begin{figure}[t]
 \center
 \resizebox{0.3\textwidth}{!}{%
  \includegraphics{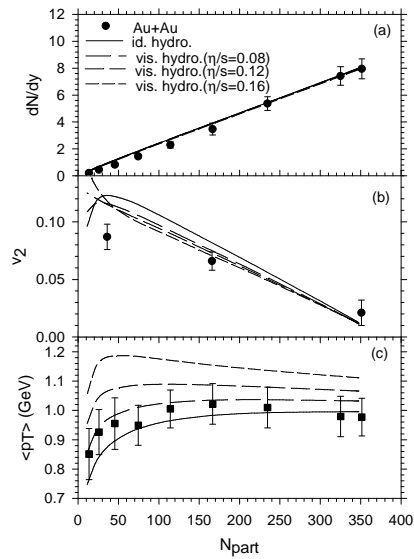}
}
\caption{STAR data on the centrality dependence of $\phi$ meson (a) multiplicity, (b) integrated $v_2$ and (c) mean $p_T$
are compared with hydrodynamical simulation of ideal and viscous fluid.}  
\label{F4}
\end{figure}

In Fig.\ref{F4}, in three panels,
STAR data \cite{Abelev:2007rw} on the centrality dependence of $\phi$ meson (a) multiplicity ($\frac{dN^\phi}{dy}$), (b) integrated $v_2$ and (c) mean $p_T$ ($<p_T^\phi>$) are shown.  
Ideal or viscous fluid, initialised to fit $\phi$ meson multiplicity in 0-5\% collisions, reproduce $\phi$ meson multiplicity   in
all the centrality ranges of collisions.  
STAR collaboration measured integrated $v_2$ only in 0-5\%, 10-40\% and 40-80\% centrality collisions. In mid-central collisions, $v_2$ is reduced by $\sim$ 10\%.
In very central collisions, elliptic flow is very small and ideal and viscous evolution produce similar flow. In peripheral collisions, both ideal and viscous evolutions overestimate the flow.  
We also observe that  the STAR data on $<p_T^\phi>$ are not explained unless $\eta/s \leq$0.12, as indicated in Fig.\ref{F4}.

\begin{figure}[t]
 \center
 \resizebox{0.3\textwidth}{!}{%
  \includegraphics{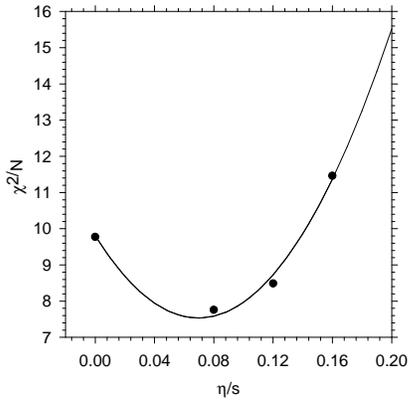}
}
\caption{Variation of $\chi^2/N$ of the combined data sets with $\eta/s$. The solid line is a parabolic fit to $\chi^2/N$.}
\label{F5}
\end{figure}

In table.\ref{table1}, we have tabulated the initial 
central energy density ($\varepsilon_i$) and temperature ($T_i$) required to fit the STAR data on $\phi$ multiplicity. The error in $\varepsilon_i$ is due to the statistical+systematic error in STAR measurement. The initial energy density of the fluid can be obtained only within $\sim$ 10-15\% accuracy. In table.\ref{table1}, we have listed the $\chi^2/N$ for the data sets analysed. Variation of $\chi^2/N$ of the combined data sets   with $\eta/s$, is shown in Fig.\ref{F5}. 
In Fig.\ref{F5}, the solid line is a parabolic fit to the combined
$\chi^2/N$, from which we estimate that the best fit to the STAR data  on the centrality dependence of $\phi$ mesons multiplicity, integrated $v_2$ and mean $p_T$ is obtained for viscosity over entropy ratio as $\eta/s=0.07 \pm 0.03$. 
We have not shown here, but hydrodynamics with $\eta/s\approx 0.07$   consistently explain transverse momentum spectra of $\phi$ mesons, along with other particles, e.g. pions, kaons. Interestingly, proton data is underestimated in the model by a factor of 2. Elliptic flow data are also reasonably well explained. In a later publication detail results will be presented.

\begin{figure}[t]
\vspace{0.3cm} 
\center
 \resizebox{0.35\textwidth}{!}{%
  \includegraphics{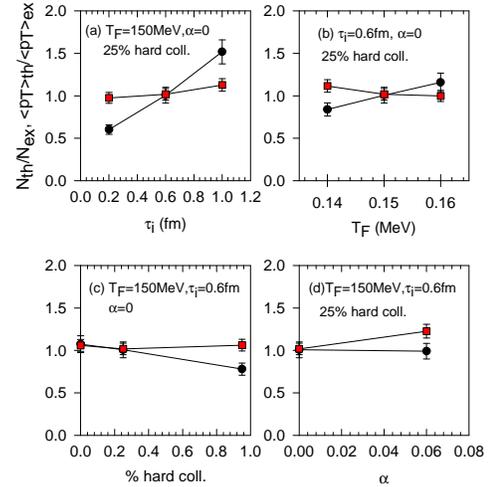}
}
\caption{(color online) The ratio of hydrodynamic predictions for $\dn$ (the black circles) and $\pt$ (the red squares), to STAR measurements in 0-5\% Au+Au collisions as a function of (a) initial time ($\tau_i$), (b) freeze-out temperature ($T_F$), (c) hard scattering contribution to initial energy density, and (d) initial transverse velocity ($v_r=tanh(\alpha r)$),      are shown. The central energy density of the fluid is $\varepsilon_i$=29.1 $GeV/fm^3$ and viscosity to entropy ratio is $\eta/s$=0.08 .}
  \label{F6}
\end{figure}

The estimate $\eta/s=0.07 \pm 0.03$ is obtained with fixed values of initial time, $\tau_i$=0.6 fm, freeze-out temperature $T_F$=150 MeV, and initial 
transverse velocity $v_r$=0. The hard scattering contribution to initial energy density was also fixed at 25\%.  
The estimate depends on the assumed initial and final conditions of the fluid.
In Fig.\ref{F6},   we have studied the dependence of $\dn$ and $\pt$ on: (a) initial time, (b) freeze-out temperature, (c) hard scattering contribution to initial energy density and (d) initial transverse velocity.  The
fluid was initialised with central energy density $\varepsilon_i$=29.1 $GeV/fm^3$,
corresponding to viscosity to entropy ratio $\eta/s$=0.08. Other conditions remaining the same, 
hydrodynamic predictions for $\dn$ and $\pt$ increases respectively by  $\sim$ 50\% and $\sim$ 10\% as   $\tau_i$ increases from 0.6 fm to 1.0 fm.
If $\tau_i$ is reduced from 0.6 to 0.2 fm,  $\dn$   decreases  by  $\sim$ 50\%, $\pt$ remains essentially unchanged. For $\tau_i$=1 fm, STAR data on $\phi$ multiplicity can be reproduced if initial energy density of the fluid is reduced by $\sim$ 40\% (we are assuming that $\dn$ dependence of initial energy density is similar at $\tau_i$=0.6 and 1 fm). However, $\sim$ 40\% reduction
in energy density will also reduce $\pt$ by $\sim$ 7\% ($\pt$ 
dependence on initial energy density is weaker than that of $\dn$) and STAR data on $\pt$ will be over predicted by $\sim$ 3\%, requiring $\sim$ 55\% less viscosity.  Arguing similarly, for $\tau_i$=0.2 fm,
to fit $\phi$ multiplicity, initial energy density is to be increased  by $\sim$ 40\%, which will also increase $\pt$   by $\sim$ 7\%, requiring $\sim$ 90\% reduction in $\eta/s$.  We estimate systematic uncertainty in $\eta/s$ due to uncertain initial time ($\tau_i=0.6 \pm 0.4$ fm) as $\sim$ 90\%.

One of the major source of uncertainty in hydrodynamic simulations is the freeze-out temperature ($T_F$). If $T_F$ is increased from 150 to 160 MeV, hydrodynamic prediction overestimate the STAR data for $\dn$ by 15\%, $\pt$ remain essentially unchanged (see Fig.\ref{F4}b).
Initial energy density of the fluid can be reduced by $\sim$10\% to fit $\phi$ multiplicity, which will simultaneously reduce
$\pt$ by $\sim$ 2 \%.  $\sim$ 2\% reduction in $\pt$ can be compensated by increasing $\eta/s$ by $\sim$ 7\%. 
If $T_F$ decreases from 150 to 140 MeV, 
hydrodynamic prediction for $\dn$ underestimate the STAR data by $\sim$ 15\% and
overestimate $\pt$ by $\sim$ 10\%. 
Initial energy density can be increased by $\sim$ 10\% to fit $\dn$, which will increase $\pt$ by $\sim$ 2\%.   $\sim$ 12\% increase in $\pt$ can not be compensated by decreasing 
$\eta/s$ and $T_F$=140 MeV will be inconsistent with STAR measurements. Indeed,
even in ideal hydrodynamics, $T_F$=140 MeV overpredict STAR data on  $\pt$. 
 We estimate the uncertainty in $\eta/s$  due to uncertain freeze-out temperature ($T_F=150 \pm 10$ MeV) as $\sim$ 100\%.
 
In Fig.\ref{F6}c, dependence of $\dn$ and $\pt$ on the hard 
scattering   contribution to initial energy density is studied. If hard scattering contribution is increased from 25\% to 95\%, hydrodynamic predictions for $\dn$ decreases by $\sim$ 20\%, but $\pt$ remain essentially unchanged. As argued previously, to fit the STAR data on $\phi$ multiplicity, initial energy density is to be increased  by $\sim$ 15\%, which will lead to increase $\pt$ by
$\sim$ 3\%. $\sim$3\% increase in $p_T$ can be compensated by reducing $\eta/s$ by $\sim$ 55\%. If hard scattering contribution decreases from 25\% to 5\%, predicted $\dn$ and $\pt$ change marginally and estimate of $\eta/s$ will not be affected. We estimate the uncertainty in $\eta/s$ due to uncertain hard scattering contribution (0-100\%)  as $\sim$55\%. 
 
Hydrodynamic predictions for of $\dn$ and $\pt$ as a function of the initial transverse velocity $v_r=tanh(\alpha r)$, for $\alpha$=0,   and 0.06 is shown in Fig.\ref{F6}d. Initial transverse velocity mainly increase high $p_T$ yield and as $\alpha$ increases from 0 to 0.06, $\dn$ remain approximately constant, but $\pt$ increases and the STAR measurement for $<p_T>$ is overpredicted by $\sim$ 20\%. Indeed, non-zero initial transverse velocity will increase $\pt$ requiring lowering of $\eta/s$.  We then estimate uncertainty in $\eta/s$ due to uncertain initial velocity as $\sim$ 100\%.
Finite accuracy in the computer code AZHYDRO-KOLKATA, also add to the uncertainty in $\eta/s$. In AZHYDRO-KOLKATA, hydrodynamic evolution is computed within $\sim$5\% accuracy, leading to  the  $\sim$7\% uncertainty in $\eta/s$. Adding all the uncertainties is quadrature, systematic uncertainty in $\eta/s$ is $\sim$ 175\%.  From the STAR data on $\phi$ meson we then estimate QGP viscosity as $\eta/s$=0.07 $\pm$ 0.03 $\pm$ 0.14, the first uncertainty is due to statistical and systematic uncertainty in STAR measurements, the second one is due to 
uncertain initial time ($\tau_i$=0.2-1.0 fm), freeze-out temperature ($T_F$=140-160 MeV), percentage of hard scattering contribution (f=0-95\%) and  initial transverse velocity ($\alpha$=0-0.6).  


It may be noted that the sources of uncertainties   considered here is not exhaustive. For example, the uncertainty in the initial   energy density may not be represented in entirety by the Glauber model, by varying only the hard scattering fraction. Color Glass condensate initial conditions, with larger initial eccentricity, may increase the range of $\eta/s$. The uncertainty in freeze-out procedure is also not entirely represented by varying freeze-out temperature only. Proper treatment of chemical freeze-out before the kinetic freeze-out, inclusion of resonances may also alter the range of uncertainty in $\eta/s$. Also, as mentioned earlier, the relaxation equation Eq.\ref{eq4}, may contain additional terms. While their contribution is expected to be smaller than the terms included, estimate of viscosity may change if a more  complete relaxation equation is used. Uncertainty in the initial shear stress tensor is also not considered here. We have used boost-invariant value as the initial shear  stress tensor. While over the time scale $\tau_\pi$, Israel-Stewart stress relax to the first order value, i.e. to the boost invariant value, it may as well be different at the initial time. Range of uncertainty in $\eta/s$ will also increase if uncertainty in initial shear stress tensor is taken into consideration.  

The estimate is obtained from experimental data, which include the effect of bulk viscosity, if there is any. We have neglected bulk viscosity. Neglect of bulk viscosity, will artificially  increase the effect of (shear) viscosity. 
In general, bulk viscosity is an order of magnitude smaller than shear viscosity. But in QGP, it is possible that near the cross-over temperature,
bulk viscosity is large  \cite{Kharzeev:2007wb,Karsch:2007jc}. Effect of bulk viscosity on particle spectra and elliptic flow is studied in \cite{Monnai:2009ad}. It appears that even if small, bulk viscosity can have visible effect on particle spectra and elliptic flow. The present estimate then must be considered as an upper bound on QGP viscosity.


\begin{thebibliography}{99}

\bibitem{Policastro:2001yc}
  G.~Policastro, D.~T.~Son and A.~O.~Starinets,
  Phys.\ Rev.\ Lett.\  {\bf 87}, 081601 (2001).
  
\bibitem{Arnold:2000dr}
  P.~Arnold, G.~D.~Moore and L.~G.~Yaffe,
  JHEP {\bf 0011}, 001 (2000),JHEP {\bf 0305}, 051 (2003).
\bibitem{Meyer:2007ic}
  H.~B.~Meyer,
  Phys.\ Rev.\  D {\bf 76}, 101701 (2007)
  [arXiv:0704.1801 [hep-lat]].
\bibitem{Nakamura:2005yf}
  A.~Nakamura and S.~Sakai,
  Nucl.\ Phys.\  A {\bf 774}, 775 (2006).
  
\bibitem{Gavin:2006xd}
  S.~Gavin and M.~Abdel-Aziz,
  Phys.\ Rev.\ Lett.\  {\bf 97}, 162302 (2006)
  [arXiv:nucl-th/0606061].
  
\bibitem{Drescher:2007cd}
  H.~J.~Drescher, A.~Dumitru, C.~Gombeaud and J.~Y.~Ollitrault,
  Phys.\ Rev.\  C {\bf 76}, 024905 (2007)
  [arXiv:0704.3553 [nucl-th]].
  
\bibitem{Lacey:2006bc}
  R.~A.~Lacey {\it et al.},
  Phys.\ Rev.\ Lett.\  {\bf 98}, 092301 (2007)
  [arXiv:nucl-ex/0609025].
  
\bibitem{Adare:2006nq}
  A.~Adare {\it et al.}  [PHENIX Collaboration],
  Phys.\ Rev.\ Lett.\  {\bf 98}, 172301 (2007)
  [arXiv:nucl-ex/0611018].

\bibitem{Luzum:2008cw}
  M.~Luzum and P.~Romatschke,
  Phys.\ Rev.\  C {\bf 78}, 034915 (2008)
  [arXiv:0804.4015 [nucl-th]].
\bibitem{Song:2008hj}
  H.~Song and U.~W.~Heinz,
  arXiv:0812.4274 [nucl-th].
  
\bibitem{Abelev:2007rw}
  B.~I.~Abelev {\it et al.}  [STAR Collaboration],
  Phys.\ Rev.\ Lett.\  {\bf 99}, 112301 (2007)
  [arXiv:nucl-ex/0703033].

\bibitem{Mohanty:2009tz}
  B.~Mohanty and N.~Xu,
  arXiv:0901.0313 [nucl-ex].

\bibitem{Alt:2008iv}
  C.~Alt {\it et al.}  [NA49 collaboration],
  Phys.\ Rev.\  C {\bf 78}, 044907 (2008)
  [arXiv:0806.1937 [nucl-ex]].

\bibitem{Koch:1986ud}
  P.~Koch, B.~Muller and J.~Rafelski,
  Phys.\ Rept.\  {\bf 142}, 167 (1986).
   
  \bibitem{strange}
  Chen J. H.  (for the STAR collaboration) J. Phys. G: Nucl. Part. Phys. 35, 104053 (2008). 
  
\bibitem{Hwa:2006vb}
  R.~C.~Hwa and C.~B.~Yang,
  arXiv:nucl-th/0602024.
  
\bibitem{Song:2008si}
  H.~Song and U.~W.~Heinz,
  Phys.\ Rev.\  C {\bf 78}, 024902 (2008)
  [arXiv:0805.1756 [nucl-th]].
  
  
  
  

\bibitem{lattice} 
Karsch F, Laermann E, Petreczky P, Stickan S and Wetzorke I, 
2001 {\it Proccedings of NIC Symposium} (Ed. H. Rollnik and D. Wolf, John 
von Neumann Institute for Computing, J\"ulich, NIC Series, vol.9, 
ISBN 3-00-009055-X, pp.173-82,2002.)

\bibitem{Cheng:2007jq}
  M.~Cheng {\it et al.},
  Phys.\ Rev.\  D {\bf 77}, 014511 (2008)
  [arXiv:0710.0354 [hep-lat]].
  
  
  
\bibitem{QGP3}
P.~F. Kolb and U. Heinz,
in {\it Quark-Gluon Plasma 3}, edited by R.~C. Hwa and 
X.-N. Wang (World Scientific, Singapore, 2004), p.~634.

\bibitem{Chaudhuri:2008sj} A.~K.~Chaudhuri,
 arXiv:0801.3180 [nucl-th].

\bibitem{note1} We have checked that with the lattice+HRG EOS, in ideal fluid dynamics, STAR measurements of $\frac{dN^\phi}{dy}$ and $<p_T^\phi>$ in 0-5\% Au+Au collisions are best explained with $T_F$=150 MeV.  


  

 





  
\bibitem{Kharzeev:2007wb}
  D.~Kharzeev and K.~Tuchin,
  JHEP {\bf 0809}, 093 (2008)
  [arXiv:0705.4280 [hep-ph]].
\bibitem{Karsch:2007jc}
  F.~Karsch, D.~Kharzeev and K.~Tuchin,
  Phys.\ Lett.\  B {\bf 663}, 217 (2008)
  [arXiv:0711.0914 [hep-ph]].

\bibitem{Monnai:2009ad}
  A.~Monnai and T.~Hirano,
  arXiv:0903.4436 [nucl-th].

\end{thebibliography}
\end{document}